# scRNA-seq of preeclamptic trophoblasts identifies *EBI3*, *COL17A1*, miR-27a-5p, and miR-193b-5p as hypoxia markers: validation of neuradapt as a superior mimetic to cobalt chloride

Evgeny Knyazev[a,b,*], Timur Kulagin[a], Ivan Antipenko[a], Alexander Tonevitsky[a,b]

[a] Faculty of Biology and Biotechnology, HSE University, Moscow, Russia

[b] Laboratory of Microfluidic Technologies for Biomedicine, Shemyakin-Ovchinnikov Institute of Bioorganic Chemistry of the Russian Academy of Sciences, Moscow, Russia

[*] Corresponding author, Faculty of Biology and Biotechnology, HSE University, 33k4 Profsoyuznaya Street, Moscow, Russia

*E-mail address*: eknyazev@hse.ru (E. Knyazev)

## Abstract

**Background**. Preeclampsia (PE) complicates 2–8% of pregnancies and involves placental hypoxia and HIF-pathway activation, especially in early-onset PE (eoPE). Chemical mimetics like cobalt(II) chloride ($CoCl_2$) and oxyquinoline derivatives model trophoblast hypoxia *in vitro*, yet their fidelity in recapitulating PE gene profiles remains unclear. Integrating patient tissue analyses with experimental models may reveal common markers and validate physiologically relevant paradigms.

**Methods**. We analyzed scRNA-seq data from 10 eoPE, 7 late-onset PE, and matched control placentas, identifying villous cytotrophoblast, syncytiotrophoblast, and extravillous trophoblast (EVT). BeWo b30 cells were treated for 24 h with $CoCl_2$ (300 μM) or the oxyquinoline derivative neuradapt (5 μM) to induce hypoxia. RNA-seq with qPCR validation and small RNA-seq quantified mRNA and microRNA changes; PROGENy inferred pathway activities.

**Results**. scRNA-seq revealed highest hypoxia activation in eoPE, with EVT showing maximum activity. Nine genes were upregulated across all trophoblast types (*EBI3, CST6, FN1, RFK, COL17A1, LDHA, PKP2, RPS4Y1, RPS26*). *In vitro*, neuradapt induced more specific hypoxia responses than $CoCl_2$ (1,284 vs. 3,032 differentially expressed genes). Critically, *EBI3, FN1,* and *COL17A1* showed concordant upregulation in tissue and neuradapt-treated cells, whereas $CoCl_2$ produced opposite patterns. MicroRNAs hsa-miR-27a-5p and hsa-miR-193b-5p were consistently elevated in both models; 3'-isoforms of hsa-miR-9-5p and hsa-miR-92b-3p were identified as hypoxia-associated.

**Conclusions**. *EBI3, COL17A1,* miR-27a-5p, and miR-193b-5p emerge as trophoblast hypoxia markers. Neuradapt (a selective HIF-prolyl hydroxylase inhibitor) provides a more physiologically relevant *in vitro* model than $CoCl_2$, recapitulating transcriptomic signatures observed in PE placentas. This integrated approach advances understanding of PE pathophysiology and therapeutic targeting.

*Keywords*

Preeclampsia, trophoblast, single-cell RNA sequencing, hypoxia, microRNA

## 1. Introduction

Preeclampsia (PE) is a serious pregnancy complication affecting 2–8% of pregnancies worldwide and representing a leading cause of maternal and perinatal morbidity and mortality [1]. This multisystem disorder originates from defective placentation characterized by inadequate extravillous trophoblast invasion and impaired spiral-artery remodelling, precipitating placental ischaemia and hypoxia [2,3]. PE presents clinical heterogeneity with distinct subtypes. Early-onset PE (eoPE; <34 weeks' gestation) features marked placental dysfunction and HIF-1α-mediated hypoxia-responsive pathway activation, while late-onset PE (loPE) shows less clear placental associations, suggesting alternative pathophysiological processes [4].

*In vitro* trophoblast models are widely employed to investigate the molecular underpinnings of PE. The BeWo choriocarcinoma cell line is widely employed as a model of trophoblasts. It is one of the few trophoblast lines that retains a broad range of placental features, including hormone secretion and the capacity for functional differentiation [5]. Specifically, the BeWo b30 subclone exhibits remarkable phenotypic plasticity, making it a versatile tool for modeling diverse trophoblast lineages. In its basal state, it forms confluent monolayers with robust tight junctions, mimicking the polarized epithelium of the villous cytotrophoblast (VCT) and the placental barrier. Upon cAMP (cyclic adenosine monophosphate) stimulation, these cells fuse to form multinucleated structures resembling the syncytiotrophoblast (SCT) [6]. Furthermore, despite their epithelial nature, BeWo b30 cells retain invasive capabilities characteristic of extravillous trophoblasts (EVT), which can be modulated by environmental cues [7]. This capacity to recapitulate key features of all three major trophoblast subpopulations—VCT, SCT, and EVT—positions the BeWo b30 clone as a unique model for identifying core molecular responses to hypoxia that are conserved across different trophoblast differentiation states. However, as a transformed cancer cell line, BeWo b30 differs from primary trophoblasts in crucial aspects, including an unlimited proliferative lifespan, altered major histocompatibility complex (MHC) expression, and a distinct metabolic profile [8]. Recognizing these limitations is essential when extrapolating *in vitro* findings to the complex, multicellular environment of the PE placenta.

To simulate hypoxic conditions in trophoblast models, chemical mimetics such as cobalt(II) chloride ($CoCl_2$) and oxyquinoline derivatives are frequently employed [9–13]. While $CoCl_2$ is the standard agent for stabilizing HIF-1α, its actions are highly pleiotropic [14]. The stabilization of HIF-1α by cobalt is classically attributed to the substitution of $Fe^{2+}$ with $Co^{2+}$ in the catalytic site of HIF prolyl hydroxylases (PHDs); however, kinetic studies suggest that $Co^{2+}$ may not effectively displace enzyme-bound $Fe^{2+}$ under physiological conditions, challenging this direct substitution model [15]. Instead, evidence indicates that $CoCl_2$ may function primarily by oxidizing ascorbate—an essential cofactor for PHD activity—rather than by direct metal displacement [16]. Additional mechanisms include direct binding to HIF-α that sterically hinders interaction with the von Hippel–Lindau protein (pVHL) [17,18] and inhibition of the asparaginyl hydroxylase FIH [19]. Furthermore, $CoCl_2$ elicits significant off-target effects: it generates reactive oxygen species and depletes intracellular glutathione [20], impairs mitochondrial membrane potential and ATP production, activates p53-dependent apoptotic pathways [21,22], induces DNA damage independently of hypoxia [23,24], and perturbs iron homeostasis by competing for metal transport proteins [25,26].

As a more specific hypoxia model, oxyquinoline derivatives (ODs) have been developed as selective PHD inhibitors. Unlike broad-spectrum iron chelators that indiscriminately sequester cellular iron—thereby impairing mitochondrial metabolism and inducing oxidative stress—advanced ODs bind directly to the $Fe^{2+}$ ion exclusively within the PHD active site [27]. Specific substituents at the 7-position of the oxyquinoline ring dictate the compound's selectivity: in the neuradapt molecule (compound 4896-3212) used in this study, a specialized 'branched tail' mimics the HIF-1α oxygen-dependent degradation domain [28,29]. This structural feature allows the inhibitor to anchor via hydrogen bonding to Asp254 in the PHD2 active site, conferring high binding specificity distinguishable from other OD variants [30]. Consequently, neuradapt selectively blocks HIF-1α hydroxylation, leading to its stabilization and activation of hypoxia-responsive pathways without the overt off-target toxicity associated with cobalt or generic chelation.

Despite the known limitations, chemical hypoxia models remain useful tools in PE research for dissecting molecular pathways, as evidenced by recent studies utilizing $CoCl_2$ and other mimetics to investigate trophoblast stress responses [31–34]. However, it remains unclear how faithfully $CoCl_2$ and ODs recapitulate the complex gene-expression profiles characteristic of PE placental tissue. While bulk RNA-seq of placental lysates has underpinned PE research, it suffers from a critical limitation: the averaging of gene expression across a heterogeneous mix of trophoblasts, immune cells, and stromal components. This 'bulk' signal obscures cell-type-specific responses and complicates comparisons with pure *in vitro* trophoblast models. Single-cell RNA sequencing (scRNA-seq) overcomes this by resolving placental heterogeneity at the individual cell level, enabling the digital isolation of specific signatures for VCT, SCT, and EVT [35,36]. By defining these lineage-specific hypoxic responses *in vivo*, scRNA-seq provides the necessary high-resolution benchmark to rigorously evaluate how faithfully $CoCl_2$ and neuradapt recapitulate the molecular landscape of PE trophoblasts.

We examined gene-expression alterations in three trophoblast subpopulations in PE using scRNA-seq data and compared these profiles with BeWo b30 cells exposed to $CoCl_2$ or neuradapt. Furthermore, to assess the translational potential of our findings, we specifically evaluated the expression of mRNAs and microRNAs previously reported to be enriched in placental extracellular vesicles (EVs) from PE pregnancies. By cross-referencing our findings with these known vesicle-associated transcripts, we aimed to determine if the molecular signatures induced by chemical hypoxia *in vitro* mirror biomarkers of the stressed placenta *in vivo*. Our ultimate goal was to contrast molecular responses of primary trophoblast subsets and hypoxia-treated cell lines, identifying robust and potentially secreted biomarkers of hypoxic perturbation in PE.

## 2. Materials and methods

### 2.1. Single-cell RNA-seq data acquisition

scRNA-seq data were obtained from the publicly available dataset generated by Admati et al. (2023) [37] on the 10x Genomics Chromium platform. The cohort comprised 10 cases of eoPE (mean delivery week 31.8 ± 3.7), 7 cases of loPE (mean delivery week 38.4 ± 1.5), and 9 gestational age-matched normotensive controls (including 3 preterm deliveries at 31.0 ± 4.3 weeks and 6 term deliveries at 39.0 ± 0.6 weeks). PE was clinically defined by the original authors according to ACOG criteria as new-onset hypertension (systolic ≥140 mmHg or diastolic ≥90 mmHg) after 20 weeks of gestation accompanied by proteinuria or evidence of end-organ dysfunction. The cohort included participants of diverse ethnicities (Caucasian-Jewish, Caucasian-Arabic, and Ethiopian). Exclusion criteria included multiple gestations, chromosomal anomalies, and chronic hypertension. Data were imported into Python (version 3.12) using scanpy (version 1.11.4).

### 2.2. Data processing and quality control

Raw UMI counts were filtered in scanpy: cells with <200 genes, genes detected in <3 cells, or >10% mitochondrial counts were removed. Doublets were excluded using scrublet. Counts were normalized to 10,000 per cell, log-transformed, and highly variable genes selected. Dimensional reduction utilized PCA, neighbor graph construction, and UMAP embedding. Subclusters from [37] were aggregated into three trophoblast categories (VCT, SCT, EVT).

### 2.3. Differential gene expression, visualization, and pathway analysis

Statitical analyses used Python with scanpy and scverse packages. Differential expression between PE and controls used sc.tl.rank_genes_groups() (Wilcoxon rank-sum test with Benjamini–Hochberg correction). Significance thresholds: $|\log_2 \text{fold change}| > 0.585$ and adjusted $p < 0.05$. Visualizations included UMAP plots and violin plots generated with scanpy and matplotlib. Gene set overlaps were visualized using jvenn [38]. Pathway activities were inferred using PROGENy (v2.1.1) for fourteen signaling pathways [39].

### 2.4. Hypoxia induction in a trophoblast cell model

The BeWo b30 subclone was selected for this study due to its unique phenotypic plasticity, which allows it to recapitulate key features of multiple trophoblast lineages including VCT-like barrier formation, SCT-like fusion potential, and EVT-like invasive capacity. This contrasts with other models such as the HTR-8/SVneo cell line, which is restricted to an EVT phenotype and lacks the epithelial integrity required to model villous trophoblast barrier functions [40]. Furthermore, unlike primary trophoblasts which exhibit significant donor variability and limited lifespan, the BeWo b30 line provides a standardized and reproducible system essential for comparative toxicological assessments of hypoxia mimetics.

BeWo b30 cells were provided by Prof. Dr. Christiane Albrecht (University of Bern, Switzerland) with permission from Prof. Dr. Alan Schwartz (Washington University in St. Louis, USA). Cells were cultured in DMEM with 10% FBS, 1% non-essential amino acids, and 1% penicillin–streptomycin (Gibco, USA) [41]. At ~80% confluence, cells were treated for 24h with 300 μM $CoCl_2$ (Sigma-Aldrich, USA) or 5 μM of the oxyquinoline derivative neuradapt (compound 4896-3212; ChemRar, Russia) to induce HIF-1α stabilization; all groups including the normoxic controls received 0.05% DMSO (vehicle control). Note: The compound neuradapt utilized here is a specific small-molecule HIF-prolyl hydroxylase inhibitor and should not be confused with the commercial dietary supplement of the same trademarked name. Control cells were maintained in a standard humidified incubator at 37°C with 21% $O_2$ and 5% $CO_2$ alongside the treatment groups. Experiments were performed using N=3 independent biological replicates for each treatment condition. Total RNA was extracted using Qiazol and miRNeasy Mini Kit with on-column DNase digestion (QIAGEN, Germany).

The concentration of neuradapt (5 μM) was selected as the minimum dose required to achieve maximal HIF-1 activation based on previous reporter assays [30]. The $CoCl_2$ concentration (300 μM) was subsequently determined via MTT assay to match the cellular viability profile of the 5 μM neuradapt treatment (Supplemental Fig. S1), ensuring that transcriptomic comparisons were performed at equi-effective doses regarding cellular stress.

## 2.5. RNA sequencing and analysis for cell model

Libraries were prepared using Illumina Stranded mRNA and NEBNext Small RNA kits. Sequencing used Illumina NextSeq 550 (75-bp single-end for mRNA; 36-bp for miRNA). Raw reads were quality-trimmed and aligned to GRCh38. miRNA isoforms were quantified with IsomiRmap. TMM normalization was applied using edgeR v4.6.3. Expression was quantified as $\log_2$-FPKM (mRNA) and RPM (miRNA). Differential expression used DESeq2 v1.48.2 with Benjamini–Hochberg correction. mRNA significance: s-values $<0.005$ and $|FC| \geq 2$; miRNA: FDR $<0.05$ and $|FC| \geq 1.5$. Functional annotation used DAVID [42]. The obtained RNA-seq results have been deposited in the Gene Expression Omnibus (GEO) under accession number GSE308908.

## 2.6. Real-time PCR analysis

RNA quality was assessed by NanoDrop and gel electrophoresis. cDNA synthesis used 1.5 μg RNA with MMLV RT kit (Evrogen, Russia), diluted 1:20 for qPCR. Reactions used 5X qPCRmix-HS SYBR kit (Evrogen), 2.5 μL cDNA, 250 nM each primer in 25 μL volumes. Cycling: 95°C/5min, then 40 cycles of 95°C/20s, 64°C/10s, 72°C/15s, followed by melting curve analysis. *ACTB* served as reference gene.

Relative expression was calculated using the $2^{-\Delta\Delta Ct}$ method from three independent biological replicates, each analyzed in technical triplicates with Student's t-test ($p < 0.05$). Primers were synthesized by Evrogen, the sequences were as follows: *EBI3* forward CCGCCTGCTCCAAACTCCA; *EBI3* reverse GGTCGGGCTTGATGATGTGC; *COL17A1* forward TACCACCAAAAGGCGGGACC; *COL17A1* reverse CCTCGTGTGCTTCCAGTTGAG; *CST6* forward TACTTCCTGACGATGGAGATGG; *CST6* reverse GAGTTCTGCCAGGGAACCAC; *HTRA4* forward CGGGGATACAGGGACCAGA; *HTRA4* reverse CAGACACTATGAACCCAGAGCC; *FSTL3* forward AACATTGACACCGCCTGGTC; *FSTL3* reverse CGTCGCACGAATCTTTGCAG; *ACTB* forward CTGGAACGGTGAAGGTGACA; *ACTB* reverse AAGGGACTTCCTGTAACAACGCA.

## 3. Results

### 3.1. Trophoblast subpopulation delineation reveals skewed cluster distributions in early- and late-onset preeclampsia

To establish a molecular benchmark for trophoblast hypoxia in PE, we first analyzed the scRNA-seq dataset [37] to delineate trophoblast subpopulations in PE. The original study profiled ten eoPE placentas (diagnosed before 34 weeks' gestation) alongside three gestationally matched controls, and seven loPE placentas with six matched normotensive controls. They defined the following trophoblast subsets: proliferating VCT (VCT_PROLIFERATING) and two postmitotic VCT clusters (VCT_1 и VCT_2); small VCT subset with an interferon-response signature (VCT_IFN); two EVT clusters, EVT_1 (high *PAPPA* and *PRG2* expression) and EVT_2 (*TNNI2* expression); classic SCT cluster with *CYP19A1* and *CGA* expression (SCT) and small placenta specific gene (PSG)-enriched SCT subset additionally expressing *PSG1*, *PSG3*, *PSG9*, *PSG11*, and *KISS1* (SCT_PSG). We reclustered these populations for eoPE and matched controls (Fig. 1A). We visualized these populations using Uniform Manifold Approximation and Projection (UMAP), a non-linear dimensionality reduction technique. In these plots, each dot represents a single cell, and cells with similar transcriptomic profiles cluster together and are color-coded by their annotated cell type to visualize discrete trophoblast subpopulations. We also aggregated EVT, SCT, and VCT clusters for combined analysis (Fig. 1B), and compared the proportional distribution of each cluster between PE and control samples (Fig. 1C). The same workflow was applied to the loPE cohort (Fig. 1D–F).

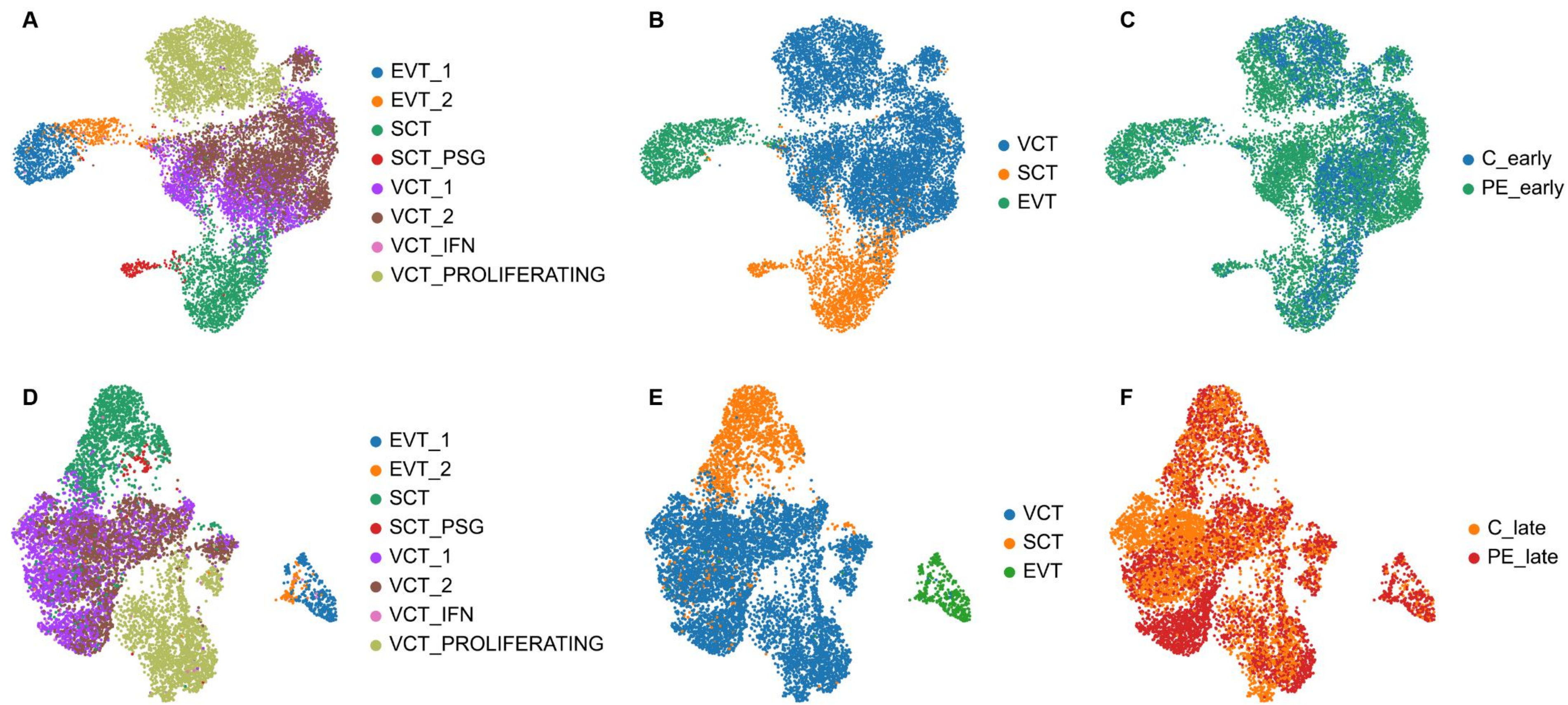

**Fig. 1**. **UMAP clustering of trophoblast subpopulations based on single-cell RNA-seq data**. **A–C.** Early-onset PE (eoPE) and matched controls. **D–F.** Late-onset PE (loPE) and matched controls. **A, D.** Annotation of trophoblast subclusters as defined by [37]. **B, E.** Aggregated trophoblast clusters. **C, F.** Relative distribution of cells between preeclampsia (PE) and control (C) samples. EVT_1, extravillous trophoblasts (EVT) with high *PAPPA* and *PRG2* expression; EVT_2, EVT with *TNNI2* expression; SCT, classic syncytiotrophoblast (SCT) fraction marked by *CYP19A1* and *CGA*; SCT_PSG, minor SCT subset additionally expressing *PSG1*, *PSG3*, *PSG9*, *PSG11*, and *KISS1*; VCT_ PROLIFERATING, proliferating villous cytotrophoblast (VCT); VCT_1 and VCT_2, postmitotic VCT clusters; VCT_IFN, VCT subset exhibiting active interferon response; C_early, controls for eoPE (delivery <34 weeks); PE_early, eoPE samples; C_late, controls for loPE (delivery ≥34 weeks); PE_late, loPE samples.

Cell counts for each trophoblast cluster and subcluster are summarized in Table 1. Although the early controls yielded the fewest total cells (3,442) and eoPE samples the most (9,867), this largely reflects the difference in cohort size (N=3 vs. N=10). When normalized per sample, cell capture rates were comparable between conditions (987 cells/sample for eoPE vs. 1,147 for early controls; 614 for loPE vs. 711 for late controls), suggesting that the disease state did not significantly compromise the technical efficiency of single-cell isolation. Notably, the VCT:SCT ratio ranged from 3.8:1 in loPE to 5:1 in eoPE, likely reflecting underrepresentation of SCT in scRNA-seq due to loss of large syncytial structures during tissue dissociation and cell capture. UMAP embedding (Fig. 1) revealed skewed cluster distributions in PE: both eoPE and loPE showed decreased VCT proportion compared to controls, while eoPE also showed reduced SCT and elevated EVT fractions suggesting altered trophoblast composition under hypoxic stress.

Table 1. Cell counts across trophoblast clusters in single-cell RNA sequencing data.

| Condition | Trophoblast subclusters as defined by [37] | | | | | | | | Aggregated trophoblast clusters | | |
|---|---|---|---|---|---|---|---|---|---|---|---|
| | VCT_1 | VCT_2 | VCT_IFN | VCT_PROL | SCT | SCT_PSG | EVT_1 | EVT_2 | VCT | SCT | EVT |
| C_early | 603 | 1203 | 27 | 895 | 632 | 26 | 31 | 25 | 2728 | 658 | 56 |
| C_late | 1194 | 1307 | 24 | 915 | 696 | 28 | 81 | 20 | 3440 | 724 | 101 |
| PE_early | 1947 | 2984 | 79 | 2229 | 1272 | 174 | 797 | 385 | 7239 | 1446 | 1182 |
| PE_late | 1037 | 1100 | 64 | 977 | 793 | 48 | 217 | 63 | 3178 | 841 | 280 |

VCT_1 and VCT_2, postmitotic villous cytotrophoblast (VCT) clusters; VCT_IFN, VCT subset exhibiting active interferon response; VCT_PROL, proliferating VCT; SCT, classic syncytiotrophoblast (SCT) fraction marked by *CYP19A1* and *CGA*; SCT_PSG, minor SCT subset additionally expressing *PSG1*, *PSG3*, *PSG9*, *PSG11*, and *KISS1*; EVT_1, extravillous trophoblasts (EVT) with high *PAPPA* and *PRG2* expression; EVT_2, EVT with *TNNI2* expression; C_early, controls for early-onset preeclampsia (delivery <34 weeks); C_late, controls for late-onset preeclampsia (delivery ≥34 weeks); PE_early, early-onset preeclampsia samples; PE_late, late-onset preeclampsia samples.

Having defined these compositional shifts, we next interrogated the functional state of these subpopulations to determine if specific lineages exhibit preferential activation of hypoxic or inflammatory signaling pathways.

### 3.2. Elevated hypoxia and inflammatory signaling in EVT and SCT

To determine if the observed compositional changes were driven by specific stress responses, we inferred the activity of 14 major signaling pathways using PROGENy [39]. This tool estimates pathway activity based on downstream target gene expression rather than the expression of signaling molecules themselves, providing a robust proxy for functional cellular states.

First, we combined all subclusters within eoPE and loPE cohorts and their respective controls to generate integrated pathway activity profiles. The results (Fig. 2A) revealed that hypoxia pathway activity was highest in eoPE samples, intermediate in loPE samples, and lowest in controls. Additionally, eoPE exhibited the strongest activation of NFκB, PI3K, TGF-β, TNF-α, and TRAIL pathways, suggesting that the hypoxic environment in early-onset disease triggers a coordinated inflammatory and apoptotic response.

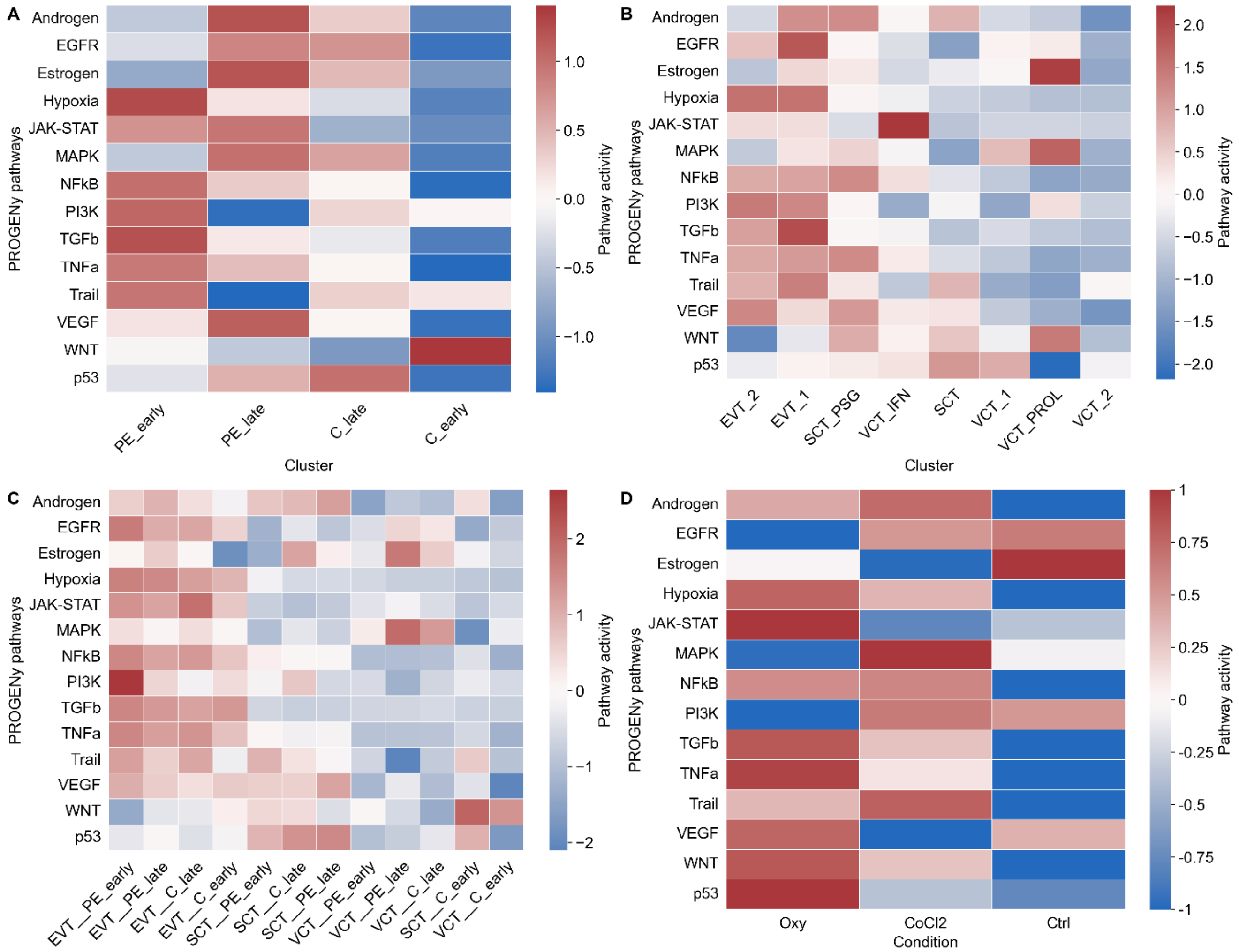

**Fig. 2. PROGENy-inferred signaling pathway activities. A**. Trophoblast clusters aggregated in sample groups. **B**. Trophoblast subclusters as defined by [37]. **C**. Aggregated trophoblast clusters. **D**. BeWo b30 in hypoxia. PE_early, early-onset preeclampsia (eoPE) samples; PE_late, late-onset preeclampsia (loPE) samples; C_late, controls for loPE (delivery ≥34 weeks); C_early, controls for eoPE (delivery <34 weeks); EVT_1, extravillous trophoblast (EVT) with high *PAPPA* and *PRG2* expression; EVT_2, EVT with *TNNI2* expression; SCT, classic syncytiotrophoblast (SCT) fraction marked by *CYP19A1* and *CGA*; SCT_PSG, minor SCT subset additionally expressing *PSG1*, *PSG3*, *PSG9*, *PSG11*, and *KISS1*; VCT_PROL, proliferating villous cytotrophoblast (VCT); VCT_1 and VCT_2, postmitotic VCT clusters; VCT_IFN, VCT subset exhibiting active interferon response; Oxy, cells treated with 5 μM oxyquinoline derivative neuradapt for 24 h at 21% $O_2$ with 0.05% DMSO vehicle; $CoCl_2$, cells treated with 300 μM cobalt(II) chloride for 24 h at 21% $O_2$ with 0.05% DMSO vehicle; Ctrl, normoxic control cells cultured for 24 h at 21% $O_2$ with 0.05% DMSO vehicle.

We next evaluated these pathway activities within specific trophoblast subclusters defined by [37], to identify which lineages are most susceptible to stress (Fig. 2B). Hypoxia signaling was most pronounced in EVT, consistent with their distance from maternal blood vessels. The SCT_PSG subset—considered the most representative syncytiotrophoblast fraction given loss of large syncytial cells during dissociation—exhibited the next highest hypoxia activity. Both EVT and SCT_PSG also showed elevated NFκB, PI3K, TGF-β, TNF-α, and TRAIL pathway activities, confirming that these functional trophoblast layers are the primary sites of molecular stress in the placenta.

For further analysis, we merged analogous subclusters into three major clusters—VCT, EVT, and SCT—and evaluated pathway activities in these aggregated clusters for PE versus controls (Fig. 2C). As before, hypoxia signaling was most pronounced in EVT, with activity highest in eoPE compared to controls. The SCT fraction in eoPE exhibited the next greatest hypoxia activity. Across all three populations, hypoxia pathway activity was elevated in PE relative to normal samples. Additionally, NFκB, PI3K, TGF-β, TNF-α, and TRAIL pathways were most active in EVT, consistent with our earlier observations.

Given these specific signaling alterations in clinical samples, we next sought to determine which chemical hypoxia induction protocol best recapitulates this hypoxic and inflammatory state in an *in vitro* trophoblast model.

### 3.3. Oxyquinoline derivative induces stronger hypoxia, TGF-β, TNF-α, and TRAIL signaling in BeWo b30 cells than $CoCl_2$

To determine if these signaling alterations observed *in vivo* could be recapitulated *in vitro*, we compared the effects of $CoCl_2$ and neuradapt on BeWo b30 cells. We induced HIF-1α stabilization using either 300 μM cobalt(II) chloride or 5 μM neuradapt, the latter providing more selective PHD inhibition. PROGENy analysis (Fig. 2D) demonstrated that while both agents elevated hypoxia pathway activity, neuradapt elicited a significantly stronger response. Crucially, when comparing these profiles to the clinical data (Fig. 2C), neuradapt treatment more faithfully reproduced the co-activation of inflammatory pathways observed in preeclamptic EVT and SCT. Specifically, neuradapt induced robust activity in TGF-β, TNF-α, and TRAIL pathways, mirroring the signature of early-onset disease. In contrast, $CoCl_2$ treatment resulted in weaker activation of these inflammatory axes and a divergent increase in PI3K signaling (which was notably decreased with neuradapt). Thus, at the pathway level, the specific PHD inhibitor provides a more accurate mimic of the inflammatory-hypoxic stress characterizing PE. However, to confirm this fidelity at the transcriptional level, we first needed to define a precise gene-set benchmark from the clinical samples.

### 3.4. Differential gene expression analysis identifies nine core hypoxia markers shared across all trophoblast clusters

To define a robust molecular PE signature that is independent of trophoblast differentiation state, we focused our differential expression analysis on the early-onset cases, which exhibited the strongest pathway-level hypoxia. While distinct lineage-specific transcriptomic alterations were observed—with 76 genes altered in EVT, 458 in SCT, and 368 in VCT—our primary goal was to identify a conserved hypoxic core that could serve as a reliable benchmark for model validation.

Intersection analysis of these lineage-specific lists revealed a highly specific set of nine genes—*EBI3*, *CST6*, *FN1*, *RFK*, *COL17A1*, *LDHA*, *PKP2*, *RPS4Y1*, and *RPS26*—that were consistently upregulated across all three trophoblast populations (Fig. 3A). This 'core signature' includes classic metabolic regulators like LDHA (lactate dehydrogenase A) and RFK (riboflavin kinase), confirming a shift toward glycolytic metabolism. Notably, it also identified EBI3 (Epstein-Barr virus-induced 3) and COL17A1 (Collagen Type XVII Alpha 1) as novel universal markers of placental stress. Conversely, six genes—*CGA*, *PLCG2*, *LY6K*, *ATP1A4*, *MTRNR2L12*, and *MTRNR2L8*—were consistently downregulated across all three trophoblast populations (Fig. 3B).

These nine universally upregulated genes represent the 'gold standard' transcriptional footprint of placental hypoxia against which our *in vitro* models must be judged. With this benchmark established, we next performed global transcriptomic profiling of our *in vitro* models to evaluate how distinct metabolic and mitochondrial programs are altered by chemical hypoxia induction.

### 3.5. Differential gene expression shows distinct and shared metabolic and mitochondrial pathway regulation upon chemical hypoxia induction

Having established the clinical benchmark, we next evaluated the global transcriptomic footprint of our two chemical models to determine their specificity. We quantified changes in protein-coding gene expression in BeWo b30 cells treated with either neuradapt or cobalt(II) chloride. Strikingly, $CoCl_2$ treatment induced a massive transcriptional perturbation, altering 3,032 genes (1,397 upregulated, 1,635 downregulated). In contrast, neuradapt treatment resulted in a more focused response, altering 1,284 genes (596 upregulated, 688 downregulated), suggesting superior specificity for hypoxia-related pathways (Fig. 3C, D).

Despite this difference in scope, intersection analysis with annotation of common genes via DAVID [42] confirmed that both agents successfully engaged the target machinery. Shared upregulated genes (193 in total) were significantly enriched in canonical hypoxia processes, including HIF-1 signaling (p=0.036), ascorbate/iron metabolism (p = 0.004), transcriptional regulation (p=0.009), and autophagy (p=0.010). Conversely, shared downregulated genes (N=281) clustered in mitochondrial function (p<0.0001), proline biosynthesis (p=0.003), cell division including cell-cycle regulation via ubiquitin-dependent proteolysis (p=0.010), fatty-acid metabolism (p=0.019), histone methylation (p=0.021), kinesin-mediated intracellular transport (p=0.032), and SLC25-mediated mitochondrial transport (p=0.041), consistent with the metabolic reprogramming known to occur in hypoxia. However, while both models broadly activated hypoxic metabolism, the critical test of their utility for PE research lies in their ability to recapitulate the specific 'gold standard' markers identified in the patient tissue.

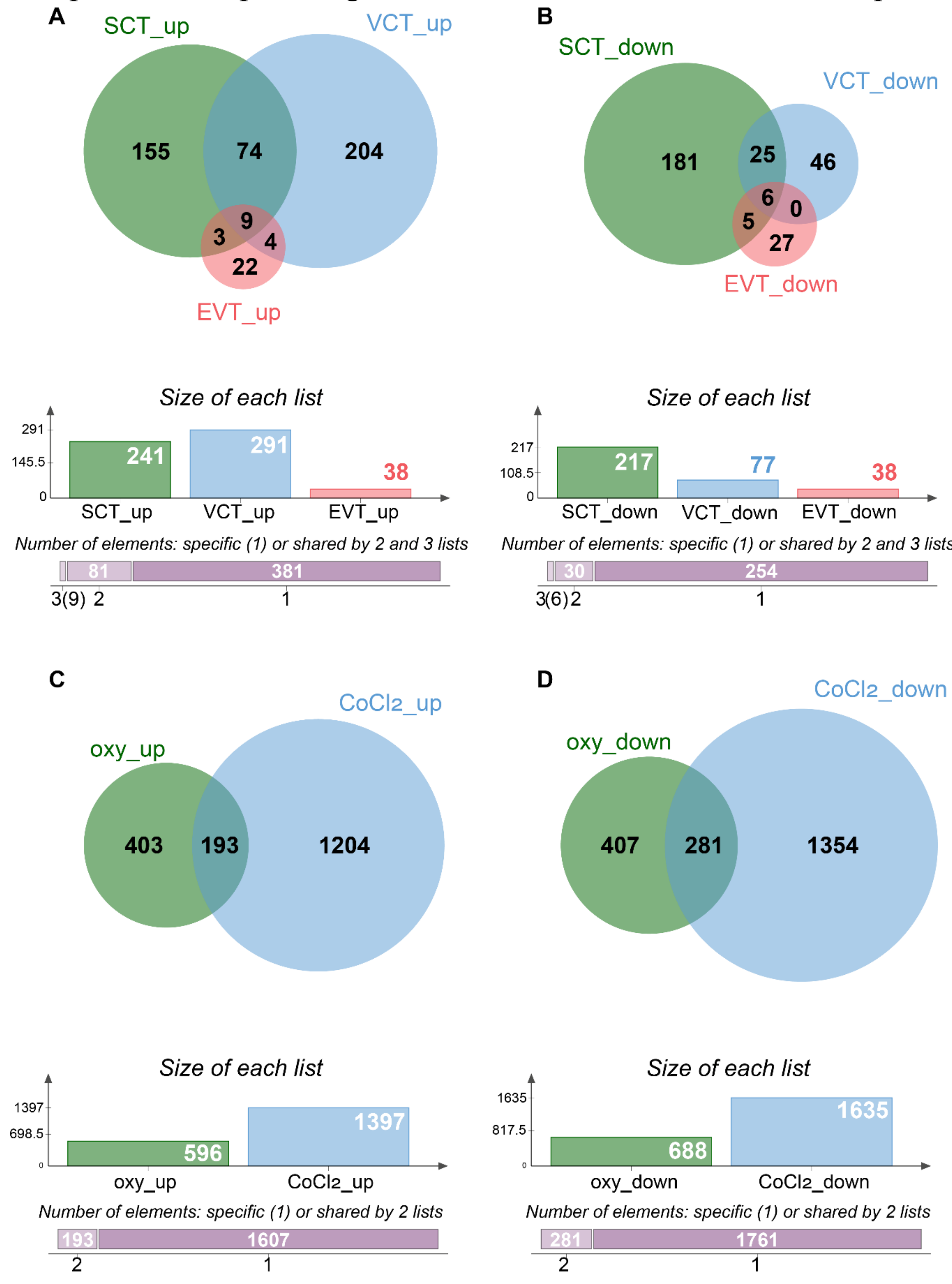

**Fig. 3**. **Differential gene-expression changes in trophoblast clusters and BeWo b30 cells**. Upregulated (A, C) and downregulated (B, D) gene sets are shown by Venn diagrams, cluster-specific counts, and numbers shared across all populations. EVT, extravillous trophoblast; SCT, syncytiotrophoblast; VCT, villous cytotrophoblast; $CoCl_2$, cobalt(II) chloride; oxy, oxyquinoline derivative neuradapt.

### 3.6. Neuradapt faithfully reproduces *EBI3* and *COL17A1* upregulation observed in preeclamptic trophoblasts

To perform this critical validation, we directly compared the expression of the universally conserved PE markers (identified in Section 3.4) between our cell models and the patient tissue. We assessed whether the nine genes upregulated across all *in vivo* trophoblast clusters were similarly regulated in BeWo b30 cells treated with $CoCl_2$ or neuradapt (Fig. 4).

This comparison revealed a striking divergence in model fidelity. While neuradapt treatment recapitulated the upregulation seen in PE for *EBI3*, *FN1*, and *COL17A1*, $CoCl_2$ exposure failed to reproduce this signature. Crucially, $CoCl_2$ actually induced an inverse regulation for these key markers—significantly repressing *COL17A1*—thereby contradicting the clinical reality.

*RFK* expression remained unchanged with either hypoxia mimetic, and *PKP2* showed a nonsignificant opposite trend relative to PE. We also examined *HTRA4* and *FSTL3*—previously reported as overexpressed in EVs from PE placentas [43]. *HTRA4* was overexpressed in SCT and VCT but downregulated in both treatments in BeWo b30 cells. *FSTL3* was overexpressed in VCT, tended to be overexpressed in EVT and SCT, was strongly induced by neuradapt, yet repressed by $CoCl_2$ in BeWo b30 (Fig. 4).

From the broader panel of 11 potential targets identified by RNA-seq (Fig. 4), we prioritized five genes for qPCR validation based on their specific relevance to model fidelity and biomarker potential. We selected *EBI3* and *COL17A1* because they exhibited the most striking divergence between $CoCl_2$ and neuradapt treatment in the sequencing data; *HTRA4* and *FSTL3* due to their established role as secreted placental vesicle markers [43]; and *CST6* as a representative shared target. We next measured the expression of this focused panel by qPCR (Fig. 5A). *HTRA4* was strongly downregulated by $CoCl_2$ and unchanged by neuradapt. *FSTL3* followed the RNA-seq trend but did not reach statistical significance. *CST6* was significantly upregulated by both $CoCl_2$ and neuradapt. *EBI3* and *COL17A1* were significantly induced by neuradapt, but showed non-significant changes in response to $CoCl_2$.

Having demonstrated that neuradapt effectively mimics the protein-coding gene signature of preeclamptic hypoxia, we finally asked whether this fidelity extends to non-coding regulatory RNAs, which are potent mediators of intercellular communication.

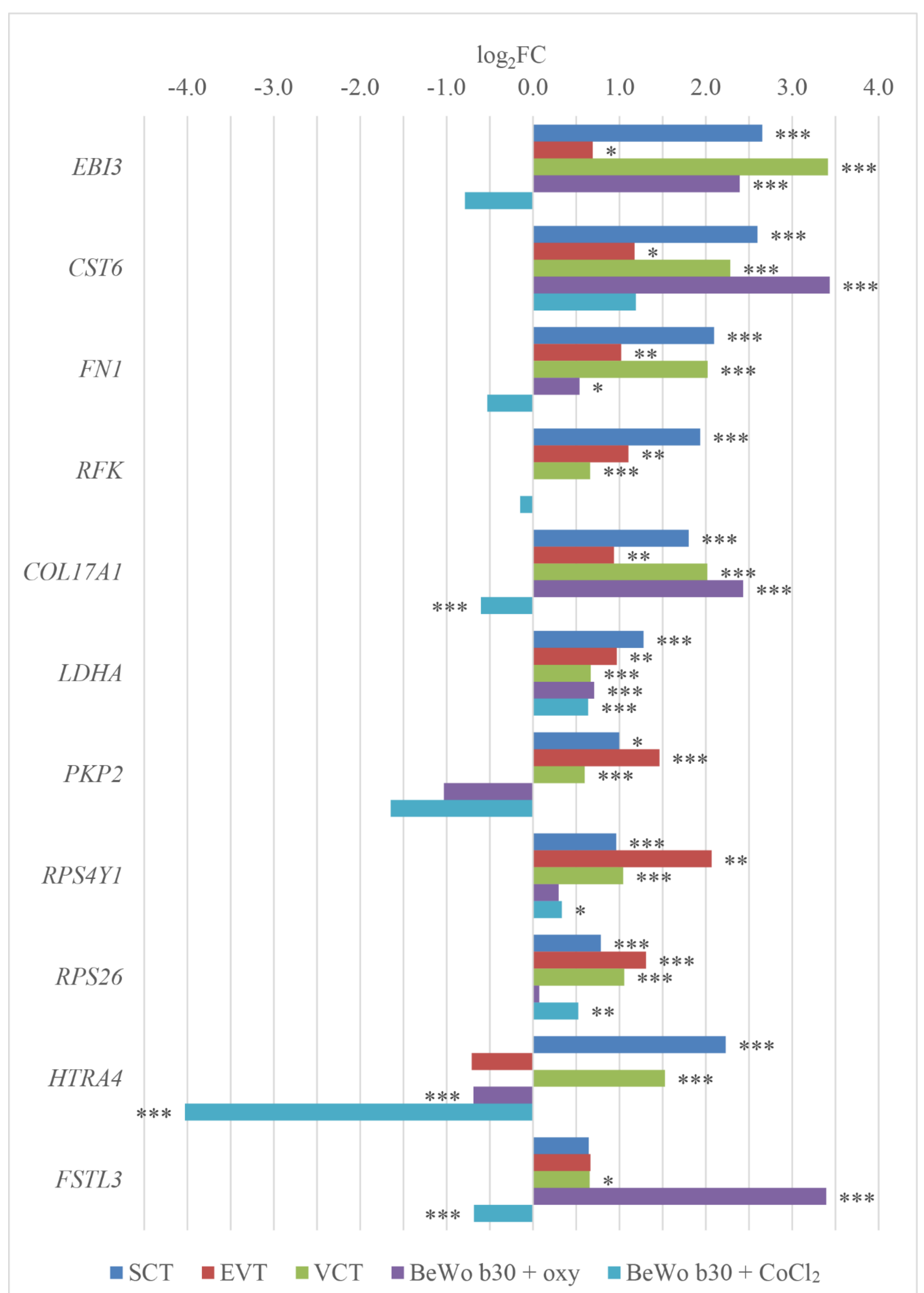

**Fig. 4. RNA-seq changes of PE markers in trophoblasts and post-hypoxia BeWo b30**. $CoCl_2$, cobalt(II) chloride; oxy, oxyquinoline derivative; VCT, villous cytotrophoblast; EVT, extravillous trophoblast; SCT, syncytiotrophoblast. * $p<0.05$; ** $p<0.01$; *** $p<0.001$.

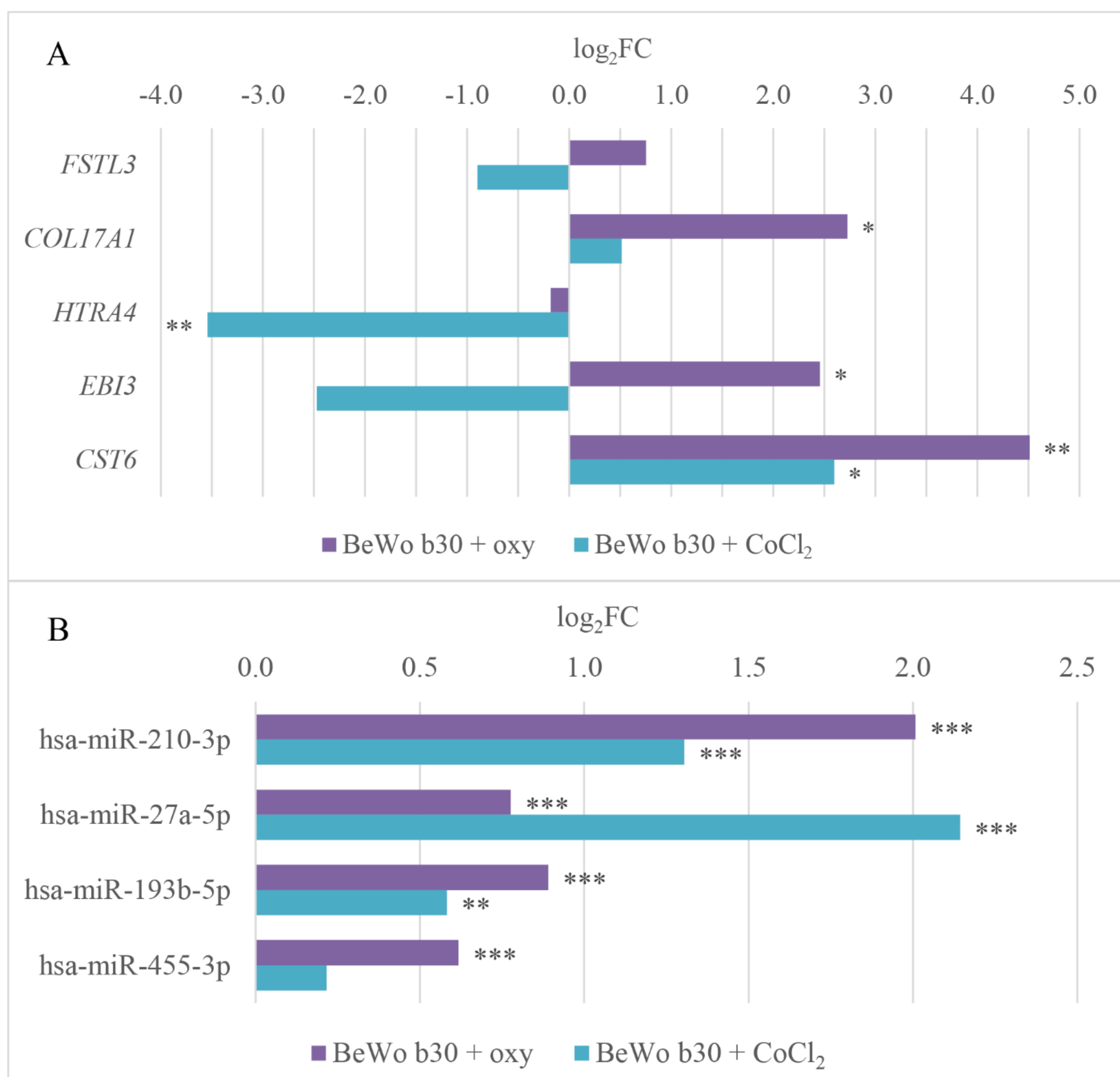

**Fig. 5. Potential preeclampsia-related hypoxia markers. A.** qPCR validation of selected genes in post-hypoxia BeWo b30. **B**. Expression changes of key microRNAs in post-hypoxia BeWo b30. $CoCl_2$, cobalt(II) chloride; oxy, oxyquinoline derivative. * $p<0.05$; ** $p<0.01$; *** $p<0.001$.

### 3.7. Chemical hypoxia induction reveals hsa-miR-210-3p, hsa-miR-27a-5p, hsa-miR-193b-5p, and 3′-isomiRs as shared PE markers

To complete our validation of the chemical hypoxia models, we investigated their ability to reproduce the non-coding RNA signatures previously identified in studies of PE placentas and circulating EVs. Since microRNAs are stable, secreted molecules that may serve as critical mediators of intercellular communication and potential circulating biomarkers [44], determining whether our *in vitro* models recapitulate their specific dysregulation is essential for translational applications. To do this, we cross-referenced our small RNA-seq data against the clinical dataset from Awoyemi et al. (2024) [45], which identified 23 miRNAs significantly upregulated in PE placentas and associated EVs.

We successfully identified a conserved 'hypoxic microRNA core' in our models. Both neuradapt and $CoCl_2$ treatment significantly upregulated hsa-miR-210-3p—the canonical 'hypoxamir'—as well as hsa-miR-27a-5p and hsa-miR-193b-5p (Fig. 5B), confirming that both agents effectively activate the hypoxia-responsive miRNA machinery. Notably, neuradapt treatment additionally elevated hsa-miR-455-3p, further broadening its coverage of the clinical signature.

Beyond canonical miRNAs, our sequencing depth allowed us to uncover a novel layer of regulation: microRNA isoforms (isomiRs). We observed significant differential regulation of non-canonical hsa-miR-9-5p isoforms, particularly in $CoCl_2$-treated cells. We observed significant upregulation of hsa-miR-9-5p isomiRs upon $CoCl_2$ treatment. Unlike canonical sequence (5′-UCUUUGGUUAUCUAGCUGUAUGA-3′), the hsa-miR-9-5p|0|-3(+2G) isoform (5′-UCUUUGGUUAUCUAGCUGUAGG-3′) showed a 2.5-fold increase ($p<0.001$), and the hsa-miR-9-5p|0|-3(+1G) isoform (5′-UCUUUGGUUAUCUAGCUGUAG-3′) increased 1.7-fold ($p=0.001$). The latter isoform also rose 1.6-fold ($p=0.002$) with neuradapt treatment. Furthermore, neuradapt exposure elevated the hsa-miR-92b-3p|0|+1 isomiR—extended by one uracil at the 3′ end—by 1.5-fold ($p=0.0001$). These findings suggest that chemical hypoxia creates a diverse landscape of regulatory RNAs that mimics the complexity of the *in vivo* environment, providing a rich resource for biomarker discovery.

## 4. Discussion

Our integrated analysis defines a precise molecular benchmark for placental hypoxia in early-onset preeclampsia (eoPE) and rigorously evaluates the fidelity of common *in vitro* models. We demonstrate that while both $CoCl_2$ and the specific PHD inhibitor neuradapt stabilize HIF-1α, they induce strikingly divergent transcriptional landscapes. Crucially, neuradapt faithfully recapitulates the specific inflammatory (TGF-β/TNF-α) and gene-expression signatures (*EBI3*, *COL17A1*) of the PE placenta, whereas $CoCl_2$ induces broad, non-specific perturbations and fails to reproduce key markers. Consequently, we propose that specific PHD inhibition in BeWo b30 cells constitutes a superior, physiologically relevant paradigm for modeling trophoblast hypoxia and discovering robust transcriptomic and isomiR biomarkers of PE.

A critical factor in the interpretation of these findings is the specific nature of the *in vitro* model employed. The BeWo b30 subclone was selected for this study because it retains a unique phenotypic plasticity, possessing the capacity for both syncytial fusion (SCT-like) and invasion (EVT-like) that other lines like HTR-8/SVneo lack [6,7,40]. Recent comparative analyses support this choice over other common lines; for instance, Li et al. (2025) highlighted that the widely used HTR-8/SVneo line is actually a heterogeneous population containing mesenchymal contaminants, lacks the critical EVT marker HLA-G, and exhibits minimal expression of the epithelial marker cytokeratin-7 [46]. In contrast, choriocarcinoma-derived lines like BeWo and JEG-3 retain the epithelial signature essential for modeling villous trophoblast biology [40] and have demonstrated superior performance in 3D vascular integration assays compared to immortalized lines [46].

However, we must rigorously acknowledge the limitations of the BeWo b30 model. The BeWo line is characterized by its inherent heterogeneity, comprising a mixed population of cells with properties resembling both syncytiotrophoblasts and extravillous trophoblasts [47]; while this plasticity is advantageous for broad screening, it introduces inherent variability that may complicate the dissection of lineage-specific mechanisms. Furthermore, global gene expression profiling reveals that BeWo cells share only a subset (~1,400) of transcripts with primary EVTs [40], and their tumorigenic origin confers a distinct metabolic profile (Warburg effect) that differs from healthy tissue [48]. Finally, our monoculture system lacks the complex stromal-immune niche that regulates invasion *in vivo*. Despite these constraints, the fact that our 'core' chemical hypoxia signature aligns with the *in vivo* single-cell data suggests that we have captured robust, fundamental stress responses that transcend the limitations of the specific cell model.

The divergent transcriptional landscapes observed in this study underscore the distinct molecular mechanisms of the two hypoxia mimetics. Cobalt chloride ($CoCl_2$), while a widely used HIF stabilizer, acts as a broad-spectrum chemical stressor with highly pleiotropic effects [14]. Although classically attributed to the substitution of $Fe^{2+}$ in the PHD catalytic site, kinetic studies suggest that $Co^{2+}$ primarily inactivates PHDs by oxidizing intracellular ascorbate [16] or depleting other cofactors [20]. Furthermore, $CoCl_2$ is known to generate reactive oxygen species, impair mitochondrial membrane potential, and induce DNA damage independently of hypoxia [21–24]. These off-target toxicities likely drive the massive, non-specific transcriptional perturbation we observed—altering over 3,000 genes including those unrelated to the HIF axis.

In contrast, oxyquinoline derivatives (ODs) represent a rational evolution in hypoxia modeling. Unlike generic chelators, advanced ODs like neuradapt are designed with a dual mechanism of action optimized through extensive structure-activity relationship (SAR) studies. As detailed by Poloznikov et al., the specific 'branched tail' at the 7-position mimics the HIF-1α oxygen-dependent degradation domain [27–30]. SAR analysis revealed that the steric bulk of this tail—specifically the substitutions on the phenyl ring and the optimized linker length—is critical for its function. It acts as a steric wedge that physically blocks the entry of the HIF peptide substrate into the PHD2 active site [27–30]. This peptide-competitive mechanism confers high specificity, distinguishing neuradapt from simple α-ketoglutarate mimetics (e.g., roxadustat, vadadustat, enarodustat) or generic iron chelators. Furthermore, safety profiling in liver-on-a-chip platforms has demonstrated that these branched-tail oxyquinolines exhibit no hepatotoxicity at concentrations up to 200 μM—40-fold higher than the effective dose used here—highlighting their superior safety profile compared to cobalt [28]. Our data confirms this mechanistic advantage: neuradapt treatment resulted in a cleaner transcriptional profile that preserved cell viability while effectively engaging the specific inflammatory programs characteristic of preeclamptic hypoxia.

Our single-cell RNA-seq analysis of PE placentas highlights both critical biological insights and inherent technical constraints. The observed skew in the VCT:SCT ratio (ranging from 3.8:1 to 5:1) confirms a significant limitation of standard scRNA-seq protocols: the systematic depletion of the large, multinucleated syncytiotrophoblast during microfluidic capture [36]. While this undersampling necessitates caution in interpreting cell abundance, the transcriptomic signatures of the surviving fractions remain robust. Single-nucleus RNA-seq (snRNA-seq) offers a potential solution by capturing syncytial nuclei with higher efficiency [49,50], though often at the cost of reduced transcript sensitivity.

Despite these capture biases, PROGENy pathway analysis successfully delineated the functional landscape of PE trophoblasts. By inferring signaling activity from downstream target expression, this method provides a robust proxy for cellular state that transcends single-gene noise [39,51]. The analysis revealed a distinct gradient of hypoxia pathway activation (eoPE > loPE > controls), providing transcriptomic validation for the central role of placental ischemia in early-onset disease [52]. Crucially, this hypoxic signature was not uniform but preferentially localized to the EVT and SCT_PSG lineages. This distribution is biologically consistent: EVT reside in the inherently hypoxic decidual niche [53], while the elevated signaling in the SCT_PSG subset likely reflects the specific metabolic stress of the functional syncytium, which is the primary site of nutrient exchange and endocrine signaling.

The coordinated activation of NFκB, PI3K, TGF-β, TNF-α, and TRAIL pathways in eoPE defines a multifaceted inflammatory state characteristic of severe placental dysfunction [4]. This signature is functionally distinct: NFκB drives the proinflammatory cytokine milieu essential for invasion [54], while concurrent PI3K activation is consistent with a compensatory survival response to hypoxic stress [55]. The robust TGF-β and TRAIL signaling observed in the EVT lineage is particularly relevant; while TRAIL is physiologically required for spiral artery remodeling, its aberrant overactivation in PE contributes to excessive trophoblast apoptosis and placental insufficiency [56,57]. The restriction of these activities to the EVT compartment underscores the specific vulnerability of the invasive lineage to ischemic insults [53].

Comparing our chemical models against this clinical benchmark reveals a crucial distinction in fidelity. While both agents stabilized HIF-1α, the specific PHD inhibitor neuradapt elicited a significantly stronger activation of the hypoxia pathway and, more importantly, faithfully reproduced the TGF-β, TNF-α, and TRAIL signature observed in preeclamptic EVTs. This suggests that neuradapt effectively mimics the inflammatory consequences of severe hypoxia without the confounding variables of generic metal toxicity [56–58]. In contrast, $CoCl_2$ treatment resulted in a divergent profile characterized by disproportionate NF-κB activation relative to other inflammatory axes, an effect attributable to cobalt-mediated oxidative stress rather than canonical hypoxic signaling [54,59]. Thus, at the pathway level, specific PHD inhibition provides a superior experimental surrogate for the inflammatory landscape of eoPE.

Transcriptomic profiling of eoPE revealed a conserved 'hypoxic core' shared across all trophoblast lineages. Despite the distinct functional roles of VCT, SCT, and EVT, nine genes—*EBI3*, *CST6*, *FN1*, *RFK*, *COL17A1*, *LDHA*, *PKP2*, *RPS4Y1*, *RPS26*—were universally upregulated, defining a fundamental molecular response to placental stress. This signature has direct translational relevance: perfusion studies confirm that *COL17A1* and *EBI3* transcripts are enriched in EVs released from PE placentas [43], validating their potential as circulating biomarkers. Importantly, our tissue-level scRNA-seq analysis also captured the upregulation of other established vesicle cargoes reported in the same study [43], including *FLNB*, *HTRA4*, and *PAPPA2* (overexpressed in SCT and VCT) as well as *INHBA* (elevated in SCT). This concordance between *in vivo* tissue transcriptomics and independent vesicular profiling underscores the robustness of these markers as indicators of placental stress that are actively secreted into the maternal circulation.

Among these core markers, *EBI3* (Epstein-Barr virus-induced 3) emerged as a critical discriminator of model fidelity. EBI3 is a subunit of the immunomodulatory cytokines IL-27 and IL-35, which regulate immune tolerance at the maternal-fetal interface [60,61]. Clinical data consistently show elevated EBI3 in decidual tissue [62] and maternal circulation [63,64] specifically in eoPE. Strikingly, while $CoCl_2$ treatment failed to induce this key marker (showing a trend toward downregulation), neuradapt treatment elicited a robust 5.5-fold upregulation, accurately mirroring the clinical phenotype. This finding is reinforced by independent evidence linking EBI3 to PE pathology: STOX1 overexpression in JEG-3 cells—a genetic model of severe PE—similarly drives a 12-fold increase in *EBI3* [65]. Furthermore, the therapeutic agent metformin has been shown to suppress EBI3 via the lncRNA-H19/miR-216-3p axis [66], suggesting that the hypoxia-induced upregulation we observed is a druggable target. To our knowledge, this is the first report of *EBI3* upregulation in a chemical hypoxia trophoblast model. Given that *EBI3* transcripts have been reported to be enriched in small EVs from PE placentas [43], our data supports the utility of the neuradapt/BeWo b30 system for identifying potential circulating biomarkers relevant to maternal immune tolerance.

The conserved hypoxic signature identified here is dominated by regulators of extracellular matrix and cell adhesion, reflecting the profound structural remodeling of the PE placenta. *COL17A1*, universally upregulated in all PE trophoblast fractions, encodes a transmembrane collagen whose ectodomain is proteolytically shed into the circulation [67]. Consistent with prior reports of elevated COL17A1 in PE tissue and its correlation with immune infiltration [68–72], our data shows that neuradapt treatment induces a robust 6.6-fold increase in *COL17A1*, whereas $CoCl_2$ fails to elicit this response. This identifies soluble COL17A1 as a promising, mechanism-based biomarker of hypoxic stress that is specifically captured by the PHD inhibitor model.

Similarly, *FN1* (Fibronectin) and its receptor *ITGA5* (an RGD-binding receptor [73]) were consistently upregulated in PE trophoblasts, indicative of altered cell-matrix interactions. The clinical relevance of this finding is underscored by reports that circulating FN1 levels are elevated in PE plasma, particularly in severe, antihypertensive-resistant cases [72,74,75]. Mechanistically, this overexpression likely involves dual epigenetic regulation: *FN1* and *ITGA5* are known to be hypomethylated in PE placentas [76], and *FN1* transcription is further enhanced by hypoxia-induced histone lactylation driven by increased glycolytic flux [77]. Our data supports this metabolic-epigenetic axis: we observed concurrent upregulation of *LDHA* (Lactate Dehydrogenase A) across all *in vivo* fractions and in both *in vitro* models, suggesting that the metabolic shift toward glycolysis precedes matrix remodeling. However, the transcriptional fidelity of the models again diverged: while neuradapt upregulated *FN1* (mimicking the disease state), $CoCl_2$ paradoxically repressed it. In contrast, *ITGA5* was robustly induced by both agents, mirroring findings in inflammatory bowel disease tissues and in Caco-2 cells under hypoxia [78], suggesting it represents a generalized epithelial response to oxygen deprivation and immune dysregulation.

Finally, *CST6* (Cystatin E/M) was reliably induced by both chemical mimetics, aligning with its reported elevation in PE tissue and maternal serum [72,79,80], and in syncytiotrophoblasts under hypoxia (1% vs. 8% $O_2$) [79]. However, unlike *COL17A1* and *EBI3*, *CST6* mRNA has not been identified in PE EVs [43], suggesting it functions primarily as a local tissue regulator rather than a secreted systemic biomarker.

Extending our validation to the non-coding transcriptome, we confirmed that both chemical mimetics successfully induce inside BeWo b30 cells a conserved 'hypoxic microRNA core' that overlaps with signatures previously reported in PE placentas and corresponding EVs. Both $CoCl_2$ and neuradapt significantly upregulated hsa-miR-210-3p—the prototypical HIF-1α–dependent 'hypoxamiR'—along with hsa-miR-27a-5p and hsa-miR-193b-5p, which have been reported as elevated in PE placental exosomes [45,81]. The validation of hsa-miR-27a-5p is particularly relevant; while its function in PE is emerging, we have previously linked it to TGF-β signaling regulation in macrophages [82], and circulating levels are known to inhibit trophoblast invasion via SMAD2 targeting [83,84]. Similarly, hsa-miR-193b-5p is a documented contributor to PE pathogenesis [85]. Notably, neuradapt treatment provided broader coverage of the clinical signature by additionally upregulating hsa-miR-455-3p, linked to PE placental tissue and gestational diabetes exosomes [86,87].

Beyond canonical sequences, our deep sequencing revealed a novel layer of regulatory complexity: the hypoxia-dependent induction of specific isomiRs. We identified significant upregulation of 3′-extended variants of hsa-miR-9-5p and hsa-miR-92b-3p. Although these modifications do not alter the seed sequence, 3′-tailoring is known to influence miRNA stability and loading into Argonaute complexes [88,89]. Notably, literature suggests that specific 3'-modifications, such as uridylation, may influence miRNA sorting and secretion [90], though the mechanisms directing isomiRs into EVs versus free circulation remain incompletely understood. The detection of these specific isoforms suggests that hypoxic stress remodels the trophoblast miRNA processing machinery, offering a new class of potential biomarkers that standard PCR assays would overlook, warranting future investigation of their secretion pathways and extracellular availability.

Collectively, our study leverages publicly available single-cell data to define a precise molecular benchmark for placental hypoxia in early-onset preeclampsia and demonstrates the superior fidelity of specific PHD inhibition over generic cobalt chelation. By rigorously cross-referencing single-cell transcriptomics with *in vitro* profiles, we identified a robust core signature—comprising *EBI3*, *COL17A1*, hsa-miR-27a-5p, hsa-miR-193b-5p, and specific hsa-miR-9-5p 3'-isomiRs—that defines the trophoblast response to hypoxic and inflammatory stress. The fact that the oxyquinoline derivative neuradapt, but not $CoCl_2$, faithfully reproduces the expression of these key immunomodulatory and structural markers in BeWo b30 cells validates it as the preferred model for dissecting the molecular mechanisms of placental dysfunction. These findings provide a validated roadmap for future biomarker discovery, emphasizing targets that are not only hypoxia-responsive but also secreted and detectable in the maternal circulation.

## 5. Limitations

Our study has several limitations that warrant consideration. First, while scRNA-seq provides high-resolution transcriptional data, analyzed clinical cohort size (N=10 eoPE, N=7 loPE) is relatively modest. Furthermore, the systematic depletion of multinucleated syncytiotrophoblasts during microfluidic capture—a known constraint of current scRNA-seq protocols—likely resulted in an incomplete characterization of the syncytial response, although we mitigated this by analyzing the surviving SCT_PSG fraction.

Second, our *in vitro* validation focused on acute (24h) chemical hypoxia. While chemical mimetics allow for precise dissection of HIF-dependent vs. off-target mechanisms, they do not perfectly replicate the complex, chronic, and fluctuating oxygen gradients (ischemia-reperfusion injury) characteristic of the PE placenta. Future studies comparing these chemical signatures directly to physical hypoxia (low oxygen levels) in trophoblast organoids or explants would further refine the model's utility.

Third, our functional validation in the BeWo b30 model relied primarily on transcriptional (RNA-seq/qPCR) and pathway-inference endpoints. While we confirmed the upregulation of secreted factors like *COL17A1* and *EBI3* at the mRNA level, confirming their release at the protein level via ELISA or Western blot is a necessary next step to establish them as clinical biomarkers. Additionally, while we identified novel isomiRs by deep sequencing, distinguishing these isoforms from canonical miRNAs using standard qPCR remains technically challenging and will require specialized probe designs for validation. Additionally, as a choriocarcinoma-derived monoculture, BeWo

b30 cells lack the critical stromal-immune interactions of the decidua; thus, our findings represent intrinsic trophoblast responses rather than the full multicellular pathology of PE.

Fourth, critically, our study did not include direct isolation, purification, or characterization of EVs, nor did we perform miRNA extraction from EV compartments. Therefore, while we identified miRNAs and mRNAs previously reported in the literature to be enriched in placental EVs from PE pregnancies, we cannot definitively conclude that the miRNAs upregulated in our BeWo b30 model are specifically packaged into or secreted via EVs. miRNA secretion is regulated by multiple mechanisms, and molecules may be released either freely or associated with EVs. Future studies employing EV isolation techniques (ultracentrifugation, size-exclusion chromatography, or immunocapture) combined with RNA extraction and qPCR or sequencing would be essential to validate whether the dysregulated miRNAs we identified are indeed preferentially sorted into EV compartments.

Finally, while our study establishes a robust transcriptomic proof-of-concept for these markers using tissue-level data, the ultimate validation of their clinical utility will require assessing their circulating levels. Future studies should focus on quantifying soluble EBI3 and COL17A1 protein and their mRNA in maternal plasma to confirm their diagnostic performance as non-invasive biomarkers of PE.

**Declaration of competing interest**

The authors declare that they have no known competing financial interests or personal relationships that could have appeared to influence the work reported in this paper.

**Funding**

The study for the sections concerning analysis of single-cell sequencing data was performed within the framework of the Basic Research Program at HSE University. The study for the sections concerning BeWo b30 cell model was funded by the Russian Science Foundation grant No. 24-14-00382, https://rscf.ru/en/project/24-14-00382/

**Declaration of generative AI and AI-assisted technologies in the writing process**

During the preparation of this work the authors used Perplexity AI tool in order to improve the readability and language of the manuscript. After using this tool, the authors reviewed and edited the content as needed and take full responsibility for the content of the published